\documentclass[a4paper]{article}
\usepackage{latexsym}
\usepackage{amsthm,amsmath,amssymb}
\usepackage[dvips]{graphicx, color}
\usepackage[title]{appendix}

\newcommand{\xdownarrow}[1]{%
  {\left\downarrow\vbox to #1{}\right.\kern-\nulldelimiterspace}
}

\usepackage{color}
\usepackage{bbm}

\newtheorem{thm}{Theorem}

\newtheorem{rem}[]{Remark}

\newcommand{\cH}{{\mathcal H}}

\newcommand{\A}{{\Bbb A}}

\newcommand{\C}{{\Bbb C}}

\newcommand{\F}{{\Bbb F}}

\renewcommand{\H}{{\Bbb H}}

\renewcommand{\O}{{\Bbb O}}

\newcommand{\Q}{{\Bbb Q}\hspace{.06em}}
\newcommand{\R}{{\Bbb R}}

\newcommand{\Z}{{\Bbb Z}}
\newcommand{\fra}{{\frak a}}

\def\O{\mathcal O}

\def\={\:=\:}  \def\+{\,+\,}

\def\a{\alpha}   \def\ba{\overline\a}    
  \def\c{\cdots}

\def\be{\begin{equation}}   \def\ee{\end{equation}}
\def\bes{\begin{equation*}}   \def\ees{\end{equation*}}
\def\ba{\begin{aligned}}   \def\ea{\end{aligned}}
\def\bc{\begin{cases}}   \def\ec{\end{cases}}
\def\bp{\begin{proof}}   \def\ep{\end{proof}}

\def\SL{\mathrm{SL}}

\newcommand{\ov}{\overline}
\def\qqan{\qquad\mathrm{and}\qquad}

\def\s{\sigma}
\def\La{\Lambda}
\def\la{\lambda}

\def\GL{\mathrm{GL}}
\def\SL{\mathrm{SL}}

\def\ov{\overline}

\def\lan{\langle}
\def\ran{\rangle}

\def\lan{\langle}
\def\ran{\rangle}

\def\bbm1{\mathbbm 1}

\def\be{\begin{equation}}   \def\ee{\end{equation}}
\def\bes{\begin{equation*}}   \def\ees{\end{equation*}}
\def\bea{\begin{equation}\begin{aligned}}   
\def\eea{\end{aligned}\end{equation}}

\def\Spec{\mathrm{Spec}}

\def\glo{\mathrm{glo}}
\def\bm{\begin{matrix}}
\def\em{\end{matrix}}
\def\bpm{\begin{pmatrix}}
\def\epm{\end{pmatrix}}

\def\rl{\rangle}

\def\diag{\mathrm{diag}}

\def\bl{\big(}
\def\br{\big)}

\begin{document}
\title{\bf   Local and Global Quantum Gates} 
\author{Lin WENG}  
\date{Graduate School of Mathematics, Kyushu University}
\maketitle
\begin{abstract}
New theories on local $p$-adic and global adelic quantum gates are developed. In particular, the corresponding universality properties are established using only finitely many local/global quantum gates.
\end{abstract}
\section{Quantum Computing}
Motivated by the theory of quantum mechanics (\cite{B}), for quantum computings, the state space of 1-qubits is defined  as a 2-dimensional $\C$-vector space with standard Hermitian inner product
\be\cH:=\Bbb C|0\rangle+\Bbb C|1\rangle\ee 
where $|0\rl=\bpm 1\\ 0\epm$ and $|1\rl=\bpm0\\ 1\epm$,
and similarly, the state space of n-qubits is defiend as the $N(=2^n)$-dimensional $\C$-vector space with standard Hermitian inner product
\be\cH^n:=\cH^{\otimes n}=\oplus_{\substack{(k_n\ldots k_2k_1)\\ k_1,k_2,\ldots, k_n\in\{0,1\}}} \C|k_n\ldots k_2k_1\ran=\Bbb C|00\cdots 0\rangle+\ldots+ \Bbb C|11\cdots 1\rangle,\ee
where $\{|k_n\cdots k_2k_1\rl:=|k_n\ran\otimes\cdots\otimes|k_2\ran\otimes|k_1\ran\ (k_i\in\{0,1\})\}$  forms an  orthonormal basis of $\cH^n$.
We call an element ${\bf x}=\sum x^{~}_{k_n\ldots k_2k_1}|k_n\ldots k_2k_1\ran\in \cH^n$  {\it normalized}
if $\sum|x^{~}_{k_n\ldots k_2k_1}|^2=1$, in which case, $|x^{~}_{k_n\ldots k_2k_1}|^2$ is called the probability for the pure qubit state $|k_n\ldots k_2k_1\rl$ to appear in ${\bf x}$, which can be observed via measurements.

Partially because normalized states should be preserved by quantum operations, which deeply rooted in the time dependence of the wave solutions for Schr\"odinger equation by the theory of quantum mechanics, unitary matrices of size $N$ are used to build up quantum gates for quantum computers when dealing with $n$ qubits. These quantum gates satisfy the so-called finite (approximate) universality properties. Consequently, each quantum circuit can be (approximately) built up from a family of finite quantum gates, consisting of Hadamard
gate, Pauli gates, Toffoli gates and their associates. For examples, we have following
quantum gates for one qubits.
\begin{enumerate}
\item [(1)] Global phase gate $$M(\a):=e^{i\a}I_2\quad \text{with}\quad I_2:=\bpm1&0\\0&1 \epm\quad \&\quad \a\in\R$$
\item [(2)] Relative phase shift $$P(\a):=\bpm 1&0\\ 0&e^{i\a}\epm\quad  \quad
 \text{with}\quad \a\in \R.$$ In particular, we set phase $\pi/4$-gate to be $S=P(\pi/2)=\bpm 1&0\\ 0&i\epm$ and 
$\pi/8$-gate to be $T=P(\pi/4)=\bpm 1&0\\ 0&e^{i\pi/4}\epm$.
\item[(3)] Pauli gates
$$\s_X=X=\bpm 0&1\\ 1&0\epm,\quad \s_Y=Y=\bpm 0&i\\ -i&0\epm,\s_Z=Z=\bpm 1&0\\ 0&-1\epm.$$
\item[(4)] Rotations with respect to $\hat x, \hat y,\hat z$ axes of the Bloch sphere
\bea
R_x(\theta)=&\cos\frac{\theta}{2}I_2-i\sin\frac{\theta}{2}X=\bpm \cos \frac{\theta}{2}&-i\sin \frac{\theta}{2}\\[0.3em]
-i\sin \frac{\theta}{2}&\cos \frac{\theta}{2}\epm\\
R_y(\theta)=&\cos\frac{\theta}{2}I_2-i\sin\frac{\theta}{2}Y=\bpm \ \cos \frac{\theta}{2}&-\sin \frac{\theta}{2}\\[0.3em]
\ \sin \frac{\theta}{2}&\cos \frac{\theta}{2}\epm\\
R_z(\theta)=&\cos\frac{\theta}{2}I_2-\ \sin\frac{\theta}{2}Z=\bpm\ \exp(\frac{\theta}{2}i)&0\\
\ 0&\exp(\frac{\theta}{2}i)\epm\\
\eea
\end{enumerate}

More generally, for 2-qubits/3-qubits, we introduce controlled-NOT/controlled-controlled NOT gate or CNOT/Toffoli gate $C/C^{(2)}$ by
\bea C|00\rl=&|00\rl,\quad C|01\rl=|01\rl,\quad C|10\rl=|11\rl,\quad C|11\rl=|10\rl.\\
C^{(2)}&|t_1,t_2,\psi\rl:=\bc|t_1,t_2,\psi\rl,&t_1t_2=0\\
|t_1,t_2,1\oplus\psi\rl,&t_1t_2=1.\ec\eea 
\begin{thm} [Approximate universality, see e.g. \cite{NC}] We have
\item[(1)] The CNOT gate together with the above 1-qubit gates is universal.
\item[(2)] The CNOT gate $C$, Hadamard gate $H$, and $\pi/8$-gate $T$ are approximately universal.
\item[(3)] The CNOT gate $C$, Hadamard gate $H$, phase gate $S$ and Toffoli gate are approximately universal.
\end{thm}

For details on fundamentals of quantum computer, quantum information and quantum programming, please refer to \cite{IK3H,NC,Y}. The reader may also find  some background materials in \cite{F}.

\section{Local $p$-Adic Quantum Computings}
Motivated by the above discussion and our own researches on distributions of zeros of non-abelian zeta functions for number fields (\cite{We2}), we now initiate an approach to what might be called the theory of local $p$-adic quantum computings (and $p$-adic quantum computers). This is based on $p$-adic probability theory of Khrennikov (\cite{K}), $p$-adic quantum mechanics  and $p$-adic string theory of, say, Freund, Witten and others.

Fix a prime integer $p$, and let $\Q_p$ be the field of $p$-adic rationals with $\Z_p$  the ring of $p$-adic integers. By definition, a  $p$-adic 1-qubit is an element of the $p$-adic quantum state space $$\cH_p:=\Q_p|0\rl+\Q_p|1\rl,$$ and a $p$-adic $n$-qubit is an element in the $p$-adic quantum state space $$\cH_p^n:=\cH_p^{\otimes n}:=\Q_p|00\cdots0\rl+\ldots+\Q_p|11\cdots 1\rl.$$ Moreover, a $p$-adic $n$-qubit ${\bf x}:=\sum x_{k_n\ldots k_2k_1}|k_n\ldots k_2k_1\rl$ is called normalized if $x^{~}_{k_n\ldots k_2k_1}\in\Z_p$ $(k_i\in\{0,1\})$.

\begin{rem} The definition of normalized states is based on the fact that $p$-adic probability is taken values in $\Z_p$ for which negative values is possible. In fact any negative natural number is considered to be of small p-adic probability. This also offers a reason why we decide to remove the constrain that 1 is the totality of probabilities since it does not make a good sense in $p$-adic setting.
\end{rem}

Ideally, $p$-adic quantum gates should preserve normalized states. Hence, it is only natural to consider the elements in the maximal compact subgroup group $\GL_N(\Z_p)$ of $\GL_N(\Q_p)$. That is to say,  an element $g\in \GL_N(\Z_p)$ defines a $p$-adic quantum gate. This appears to be perfectly compatible with  nowadays mathematics: With respect to the structure
of the general linear group $\GL_N$ over the adeilc ring $\A=\A_F$ associated to a global algebraic number field $F$, it is central for us to introduce its canonical maximal open compact subgroup $K:=\prod_{\frak p:\text{finite}}K_{\frak p}\times\prod_{\s|\infty}K_\s$ where for each finite place $\frak p$ of $F$, $K_{\frak p}=\GL_N(\O_{\frak p})$ is the general linear group over the local ring $\O_{\frak p}$ of $\frak p$-adic integers, and for each infinite place $\s$ of $F$, when $\s$ is real, resp. complex, we set $K_\s=O_N$, resp. $U_N$, the orthogonal group, resp., the unitary group,  of size $N$.

Note that $GL_N(\Z)$ is finitely generated and moreover is dense in $\{\pm 1\}\times \SL_N(\Z_p)$ (with respect to the topology induced from the $p$-adic topology), it is in principal not very difficult but rather fundamental to establish the $p$-adic approximate universality theorem for $p$-adic quantum gates. Before explain the details, let us make some preparations.

For an element $a\in \Q_p$, set 
\be
E(a):=\bpm a&0\\ 0&1\epm\qquad U(a):=\bpm 1&a\\ 0&1\epm \qqan L(a):=\bpm 1&0\\ a&1\epm\ee
In addition, as usual, 
when $p\not=2$, we introduce the $p$-adic exponential $\exp_{p}(pa):=\sum_{n\geq 0}\frac{p^na^n}{n!}$ as the isomorphism of the additive group $:p\Z_p$ onto the multiplicative group $1+p\Z_p$ such that $|\exp_p(pa)-\exp_p(pb)|=|p||a-b|$, whose reciprocal isomoriphism is given by the $p$-adic logarithmic function
$\log_p(1+pa)=\sum_{n\geq 1}(-1)^{n-1}\frac{p^na^n}{n}.$\footnote{when $p=2$, we instead consider the isomorphism $\exp_2(2^2a)=\sum_{n\geq 0}\frac{2^{2n}a^n}{n!}$, an isomorphism
of the additive group $2^2\Z_2$ onto the mulriplicative group $1+2^2\Z_2$ such that $|\exp_2(2^2a)-\exp_2(2^2b)|=|2^2||a-b|$ whosereciprocal isomorphism is given by the 2-adic
logarithmic function $\log_2(1+2^2a)=\sum_{n\geq 0}=(-1)^{n-1}\frac{2^{2n}a^n}{n}$.} Fix also  a primitive $(p-1)$-th root $\zeta_p$ of unity in $\Z_p$.

\begin{thm} [$p$-Adic Approximate Universality] The following set of $p$-adic quantum gates
is approximately  universal:
\begin{enumerate}
\item [(1)] if $p\not=2$,
$$\left\{M_\zeta:=E(\zeta_p),\ E(\exp_p(p)),\ U(1),\ L(1)
\right\}$$
or
$$\left\{M_\zeta:=E(\zeta_p),\ P_{1+p}:=E(1+p), \ P_+:=U(1),\ P_-:=L(1)
\right\}$$
\item [(2)] if $p=2$,
$$\left\{M_{-1}:=E(-1),\ P_{2^2}:=E(\exp_2(2^2)),\ P_+:=U(1),\ P_-:=L(1)
\right\}$$
or
$$\left\{M_{-1}:=E(-1),\ P_{1+2^2}:=E(1+2^2),\ P_+:=U(1),\ P_-:=L(1)
\right\}$$
\end{enumerate}
\end{thm}
\bp
Since the proof for $p=2$ is similar, we here only give a proof for $p\not=2$ and briefly
sketch how modifications can be made to apply the same argument to  the case $p=2$. Let us first consider the case $N=2$. Obviously, $GL_s(\Z_p)$ admits a natural semi-product  decomposition
\be
\GL_2(\Z_p)=E(\Z_p^*)\rtimes\SL_2(\Z_p).
\ee
Hence it suffices to find the topological generators of the $p$-adic unit group $\Z_p^*$ and $\SL_2(\Z_p)$. First we deal with $\Z_p^*$ by considering the canonical quotient map
$\Z_p\mapsto \Z_p/p\Z_/p=\F_p$. Obviously, $\F_p^*=\lan\zeta_p\ran$ as a cyclic group of order $p-1$, and an element $x\in Z_p$ belongs to $\Z_p^*$ if and only if \be
z=\zeta_p^{n(z)}\bl 1+pb(z)\br=\zeta_p^{n(z)}\exp\bl\exp_p(p\a(z)\br\qquad\exists \a(z):=\frac{\log_p(1+pb(z)}{p}\in \Z_p.
\ee
Therefore, if we set $\a_n(z):=\sum_{k=0}^{n-1}\la_k(z)p^k$ be the $(n-1)$-th truncated sum of teh $p$-adic expansion of $\a(z)$, then
$$
1+pb(z)=\exp_\bl(p\a(z)\br=\lim_{n\to\infty}\exp_p\bl p\a_n(z)\br=\lim_{n\to\infty}\exp_p(p)^{\a_n(z)}.
$$
Hence
$$
z=\zeta_p^{n(z)}\cdot \lim_{n\to\infty}\exp_p(p)^{\a_n(z)}.
$$
In other words, $\Z_p^*$ is topologically generated by $\{\zeta_p,\exp_p(p)\}$.

In addition, using the $p$-adic logarithmic function $\log_p(1+pa)$, any element $x=1+pc$ of $1+p\Z_p$ can be written uniquely as
$$(1+p)^{\a(x)}=\sum_{k\geq 0}{\binom{\a(x)}{k}}p^k=\exp_p(\a(x)\log_p(1+p))\quad\left(\exists\a(x):=\frac{\log_p(1+pc)}{\log_p(1+p)}\right).$$ Thus for a sequence $\{\a_n(x)\}_n\subset\Z_{\geq 0}$ satisfying $\a(x)=\lim_{n\to\infty}\a_n(x)$, we have
$$x=\lim_{n\to\infty}(1+p)^{\a_n(x)}.$$
This implies that $\{\zeta_p,1+p\}$ topologically generates $\Z_p^*$ as well.

When $p=2$, we have $\Z_2^*=\{-1,1\}\times (1+2^2\Z_2)$. Hence the above arguments works  well as claimed.

From the discussions above, to complete the proof of this theorem for $N=2$, it suffices to show that $\SL_2(\Z_P)$ is topologically generated by $P_+$ and $P_-$ together with $E(\Z_p^*)$.

For this, let $s=\bpm a&b\\ c&d\epm\in \SL_2(\Z_p)$. Following \cite{M}, we note that, if $c$ is a $p$-adic unit, with $ad-bc=1$, or the same $b=(ad-1)c^{-1}$,
\bea
\bpm a&b\\ c&d\epm=&\bpm 1&-(1-a)c^{-1}\\ 0&1\epm\bpm 1&0\\c&1\epm\bpm 1&-(1-d)c^{-1}\\ 0&1\epm\\
\bpm a&c\\ b&d\epm=&\bpm 1&0\\-(1-a)c^{-1}&1\epm\bpm 1&c\\0&1\epm\bpm 1&0\\
-(1-d)c^{-1}&1\epm.
\eea
Each of the factors on the right hand sides of both relations can be topologically generated by $P_+$ and $P_-$.
Hence we may assume that  $c$ is not a $p$-adic unit. This implies that $a$ is a $p$-adic unit since $ad=1+bc$. Hence we may instead consider the matrix
$$\bpm c&d\\-a&-b\epm= \bpm 0&1\\ -1&0\epm\bpm a&b\\ c&d\epm$$
and apply the above argument. This then complete the proof for the case $N=2$ since 
$$
 \bpm 0&1\\ -1&0\epm=P_+P_-^{-1}P_+.$$

To deal with general $N$, it suffices to use the Schmit normal form for  matrices
$A=(a_{ij})\in M_N(\Z_p)$. In fact, it is a standard fact that there exist two elements $L,R\in \GL_N(\Z_p)$ such that $LAR=\diag(d_1,d_2,\ldots,d_N)$, where $d_1,d_2,\ldots, d_N$, the so-called elementary divisors  of $A$, satisfy the condition that
$d_1|d_2|\ldots|d_N$ and $d_i=\Delta_i/\Delta_{i-1}\ (1\leq i\leq N)$,
where $\Delta_0=1$ and $\Delta_i\ (1\leq i\leq N)$ denotes the gcd of all $i\times i$-minors of $M$. Furthermore, $L$ and $R$ are products of the elementary matrices of the following three types: 
\bea
T_{ij}(b)=&I_N+bE_{ij}\ (1\leq i\not=j\leq N, b\in \Z_p),\\
P_{ij}:=&I_N-E_{ii}-E_{jj}+E_{ij}+E_{ji}(1\leq i\not=j\leq N),\\
 D_i(u):=&I_N-(1-u)E_{ii}\ (1\leq i\leq N, u\in \Z_p^*).
\eea
Here  $I_N$ denotes the identity matrix of size $N$ and $E_{ij}\in M_N(\Z_p)$ denotes the matrix whose $(k,l)$-entry is 1 if $(k,l)=(i,j)$, 0 otherwise. 
\ep

\section{Global Adelic Quantum Computings}

Quantum computers and $p$-adic quantum computers should be viewed as local versions of
more global quantum computers. For this reason,  we sometimes call current quantum computers analytic quantum computers,  while saving the terminology of quantum computer for both analytic quantum and $p$-adic quantum computers. 

Recall that in mathematics, naturally associated to global fields $F$ are local fields first, both archimedean $F_\s$  and non-archimedian  $(F_v,\O_v,k_v)$, and then the global adelic ring 
$\A_F$, defined as the restricted product of $F_v$ and $F_\s$ with respect to $\prod_v\O_v$. In parallel, for the theory of quantum computers, there should be one for what might be called global or adelic quantum computers. 

Our first task is to understand what should be the state space $\cH_\A$. It is only natural to take it to be $\A^N$ or more generally $\oplus_{i\in I}\A$ where $I$ is a countable index set. For simplicity, in the sequel, we only work over finite 'dimensional' state space, namely, assuming
that  $\#I=N<+\infty$. A state vector $x=(x_v;x_\s)\in  \cH_\A$ is called normalized if for all finite places $v$, resp. infinite places $\s$ of $F$, $x_v$, resp. $x_\s$ are locally normalized. Consequently, adelic quantum gates should be associated to the elements of a maximal compact subgroup $K=\prod_v K_v\times\prod_\s K_\s$ of $\GL_N(\A)$ with $K_v=\GL_N(\O_v)$ and $K_\s$ are either orthogonal group $O_N$ or unitary group $U_N$ depending whether $\s$ is real or complex. 

To be more precise, 2-dimensional adelic state space consisting of  adelic one qubit  is defined to be $\cH_{\A,2}:=\A|0\ran+\A|1\ran,$ and more generally, $N=(2^n)$-dimensional adelic state space consisting of  adelic $n$ qubit  is defined to be $$\cH_{\A}^n:=\bigoplus_{\substack{(k_n\ldots k_2k_1)\\ k_1,k_2,\ldots, k_n\in\{0,1\}}}\A|k_n\ldots k_2k_1\ran.$$ An adelic $n$-qubit
$$a=\sum_{\substack{(k_n\ldots k_2k_1)\\ k_1,k_2,\ldots, k_n\in\{0,1\}}}a_{k_n\ldots k_2k_1}|k_n\ldots k_2k_1\ran\in \cH_{\A,N}$$ is called normalized if
 for each $(k_n\ldots k_2k_1)\in\{(0\ldots 0 0),\ldots(1\ldots 1 1)\}$,  $a_{k_n\ldots k_2k_1}$ $=(a_{k_n\ldots k_2k_1,v};a_{k_n\ldots k_2k_1,\s})\in \A$, and 
for each finite place $v$, $a_{k_n\ldots k_2k_1,v}\in \O_v$ while for each infinite place $\s$,
$$\sum_{\substack{(k_n\ldots k_2k_1)\\ k_1,k_2,\ldots, k_n\in\{0,1\}}}|a_{k_n\ldots k_2k_1,\s}|^2=1.$$
Furthermore, among all elements $g=(g_v;g_\s)\in \GL_N(\A)$, adelic quantum gates should be build up merely from the elements of its maximal compact subgroup $$K_N:=\prod_{v:\mathrm{finite}}\GL_N(\O_v)\times\prod_{\s:F\hookrightarrow \R}O_N\times\prod_{\tau:F\hookrightarrow \C}U_N.$$

The first difficulty faced by adopting this approach to adelic quantum gates is that, for each such a adelic quantum gate, there are infinitely many operations which should be proceeded. To 
understand this, we may recall the so-called strong approximation theorem for algebraic groups. 

Let $G$ be an algebraic group over a global field $F$. Within the adelic ring $\A$ of $F$, for a 
non-empty finite set $S$ of places of $F$, define $\A^S$ to be the ring of $S$-adeles and 
$\A_S$ to be the product of $K_v$'s for all $v\in S$. Obviously, via the diagonal embeddings, $G(F)$ can be viewed as a subgroups of both $G(\A^s)$ and $G(\A_S)$. Then the weak approximation is a property that  $G(F)$ is dense in $G(\A_S)$, while the strong approximation is a property that  $G(F)$ is dense in $G(\A^S)$. This later property is equivalent to that
$G(F)G(\A_S)$ is dense in $G(\A)$. If so then the approximate university can be answered using elements of $G(F)$, since $G(\A_S)$  always satisfies approximate university for the reason that $S$ is finite.

For general algebraic groups, the strong approximation is not satisfied. However, when $G$ is a semi-simple and simply-connected,  the strong approximation holds, established by Kneser 
(\cite{Kn}) and Platonov (\cite{P,PR}), resp. Margulis (\cite{M}) and Prasad (\cite{Pr}), when $F$ is a number field, resp. a function field over a finite field.

Related to this, one may wonder how the classical reduction theory enters into the picture.
In case $\SL_2$, we may consider the quotient $\SL_2(F)\backslash\SL_2(\A)/K_2$. This is studied intensively in (\cite{We}) using stability, as an integrated part of rank two non-abelian zeta function of $F$. In terms of integral model to be discussed below, this means that the global quantum gates are build up from $\SL(\O_F\oplus\fra)$, which enjoys the university property since $\SL(\O_F\oplus\fra)$ is finitely generated (See e.g. \cite{EGM}), and
admits a natural action on $\cH^{r_1}\times\H^{r_2}$ where $\cH$, resp. $\H$, denotes the hyperbolic upper half plane resp. the 3-dimensional hyperbolic upper half space. In this way, then geometry is naturally involved.\footnote{For example, we may viewed both upper half plane
$\cH=\R|0\ran+\R_{\geq 0}|1\ran$ and 3-dimensional hyperbolic upper half-space 
$\H=\R+\R i+\R_{\geq 0}j$
as  topological subspaces in $\cH=\C|0\ran+\C|1\ran$. Despite the fact they are not vector subspaces, the action on these subspaces by global quantum gates still make perfect sense.}

To avoid this, we may use the adelic topology
involved to do the approximation. The ideal situation then should be for each adelic quantum gate, there should be only finitely many places which are not trivial. For our own use, we call such adelic quantum gates one of finite type.
Put this in an equivalent way, in terms of adelic quantum state vectors, for the difference between the input and the output adelic quantum state vectors, there is a finite set $S$ of places, including possibly parts of infinite places, such that
for $v\not\in S$, the $v$-component of the adelic quantum gates is trivial, i.e. degenerates to the identity operators. This then would mean that the difference between the input and output  state vectors should be the same for almost all but finitely many components. In other words, there should be  a global integral structure involved for adelic quantum calculations. To understand this, we supply the following discussion on global quantum computings.

\subsection{Integral Structures}

Even this first level consideration above would certainly lead to a nice and rich theory of adelic quantum computers, 
it may prove to be too complicated for us human beings to achieve in  even a distance future.
Accordingly, as a more realistic goal, we may instead work over a global integral version of this adelic theory. That is to say, we only work over finite rank $\O_F$-lattices $\La=(P,\rho)$, or the same, the metrized locally free sheaves of finite rank over arithmetic curves $\ov X=\ov{\Spec\O_F}$. Here $P$ denotes a projective $\O_F$-module of finite rank $N=2^n$ over the ring of integers $\O_F$ of $F$, and $\rho$ denotes a compatible metric on  the Minkowski space
\bea
P\simeq &\O_F^{N-1}\oplus\fra\hookrightarrow \oplus_{\substack{(k_n\ldots k_2k_1)\\ k_1,k_2,\ldots, k_n\in\{0,1\}}} F|k_n\ldots k_2k_1\ran\\
\hookrightarrow& \oplus_{\substack{(k_n\ldots k_2k_1)\\ k_1,k_2,\ldots, k_n\in\{0,1\}}} F_\infty|k_n\ldots k_2k_1\ran\\
=&\oplus_{\substack{(k_n\ldots k_2k_1)\\ k_1,k_2,\ldots, k_n\in\{0,1\}}} (\R^{r_1}\times\C^{r_2})|k_n\ldots k_2k_1\ran\\
=:&\cH_{\glo}^n.\eea
Here $\fra$ denotes a fractional ideal of $F$, and $r_1$, resp. $r_2$ denotes the number of real, resp. complex places of $F$\footnote{Indeed, since $\O_F$ is a Dedekind domain, it is well known that for each fixed projective $\O_F$-module $P$ of rank $N$, there exists a fractional ideal $\fra(P)$ of $F$ such that as $\O_F$-modules, $P\simeq \O_F^{N-1}\oplus\fra(P).$ For details, see e.g. \cite{We}.}.
A element $x\in\cH_{\glo}$ is called a state vector and a state vector $x$  is called normalized if $x$ is in an Minkowski image of $P$.
In this way, we are lead to build up global quantum gates through elements of $\GL_N(\O_F)$. Since $\GL_N(\O_F)$ is known to be of finitely generated, global (quantum) gates satisfy universality property. 

To simplify our presentation, let us assume that $F=\Q$, the field of rationals. 
Then the fact that global quantum gates enjoy the universality property can be justified as follows. For $N=2$, $\GL_2(\Z)$ is well known to be generated by
$$\left\{Z=\bpm 1&0\\ 0&-1\epm,\ X=\bpm 0&1\\1&0\epm,\ P=\bpm 1&1\\ 0&1\epm\right\}.$$
More generally, thanks to Hua and Reiner (\cite{HR}), we know that $\GL_N(\Z)$ is generated by two elements. To be more precise, for any positive integer $m$, the general linear group $\GL_m(\Z)$ of size $m$
 is generated by $X$ and $P$ when $m$ is even and by $-X$ and $P$ when $m$ is odd. Here
$$
X=\bpm 0&0&0&\cdots&0&1\\
1&0&0&\cdots&0&0\\
0&1&0&\cdots&0&0\\
\cdots&\cdots&\cdots&\cdots&\cdots&\cdots\\
0&0&0&\cdots&0&0\\
0&0&0&\cdots&1&0\epm, \ \ P=\bpm 1&1&0&\cdots&0&0\\
0&1&0&\cdots&0&0\\
0&0&1&\cdots&0&0\\
\cdots&\cdots&\cdots&\cdots&\cdots&\cdots\\
0&0&0&\cdots&1&0\\
0&0&0&\cdots&0&1\epm.
$$
In terms of lattices, $X$ corresponds to the state swap, and $P$ corresponds to the state shift. Since these gates can be realized classically, we may use classical computer to help us to understand this type of global quantum computings. In particular, these global  gates are   adelic quantum gates of finite type.

Similarly, then the reduction theory for reductive groups naturally enters into the picture.
Classically, the theory is developed using Siegel sets. But the classical approach is not neat since nowhere precisely estimations are needed. In contrast, a new very much powerful approach is adopted in (\cite{We} working over number fields, \cite{WZ,WZ2} working over function fields) using stability,  From this new approach, the involved fundamental domains and their associated cusps are classified and partitioned according to various levels the parabolic subgroups of the groups involved. Even in this sense, we hope that our discussions here on local and global quantum gates certainly opens a narrow door to these wonderful parallel worlds with vast fertile lands and rich structures, and that the studies on what might be called local $p$-adic quantum computers and global adelic quantum computers would become more and more attractive.
\vskip 0.30cm
{\bf Acknowledgement}. We would like to thank Zhan SHI for his discussions, during which a slip in the first version was detected.

\vskip 13.0cm
\noindent
Lin WENG\\
Graduate School of Mathematics,\\
Kyushu University,\\
Fukuoka 819-0395,\\
JAPAN\\
E-Mail: weng@math.kyushu-u.ac.jp


\begin{thebibliography}{50}
\bibitem{B} D. Bohm, {\it Quantum Theory},  Prentice-Hall, Inc. 1951
\bibitem{EGM} J. Elstrodt, F. Grunewald, J.Mennicke, {\it Groups Acting on Hyperbolic Spaces: Harmonic Analysis and Number Theory}, Springer 1997
\bibitem{F} K. Fujii, {\it Amazing Quantum Computer}, (in Japanese) Iwanami Shoten, 2019

\bibitem{HR} L.K. Hua, I. Reiner,  Automorphisms of the unimodular group. Trans. Amer. Math. Soc. 71 (1951), 331-348.
\bibitem{IK3H} S. Ishizaka, T. Ogawa, R. Kawauchi, G. Kimura, M. Hayashi, {\it Introduction to Quantum Information Science}, (in Japanese) Kyoritsu Shuppan, 2012
\bibitem{K}  A. Khrennikov, Interpretations of probability and their $p$-adic extensions, (in Russian) Teor. Veroyatnost. i Primenen. 46 (2001), no. 2, 311-325; translation in Theory Probab. Appl. 46 (2003), no. 2, 256-273

\bibitem{Kn} M. Kneser, Strong approximation, {\it Algebraic Groups and Discontinuous Subgroups} (Proc. Sympos. Pure Math., Boulder, Colo., 1965), Providence, R.I.: American Mathematical Society, pp. 187-196
\bibitem{Ma} G. Margulis, Cobounded subgroups in algebraic groups over local fields, Akademija Nauk SSSR. Funkcional'nyi Analiz i ego Prilozenija, 11 (2): 45-57
\bibitem{M} T. Mounkoro, Some subgroups of the general linear group of order two over the ring of $p$-adic integers.  $p$-Adic Numbers Ultrametric Anal. Appl. 6 (2014), no. 3, 219-234. 






\bibitem{NC} M. Nielsen and I. Chuang, {\it Quantum Computation and Quantum Information}, 10th Anniversary Edition, Cambridge University Press, 2010

\bibitem{P} V. Platonov, The problem of strong approximation and the Kneser-Tits hypothesis for algebraic groups, Izvestiya Akademii Nauk SSSR. Seriya Matematicheskaya 33, 1211-1219

\bibitem{PR} V. Platonov, A. Rapinchuk, {\it  Algebraic groups and number theory}. (Translated from the 1991 Russian original by Rachel Rowen.), Pure and Applied Mathematics, 139, Boston, MA: Academic Press
\bibitem{Pr} G. Prasad, Strong approximation for semi-simple groups over function fields, Annals of Mathematics, Second Series, 105 (3): 553-572

\bibitem{We} L. WENG, {\it Zeta Functions of Reduction Groups and Their Zeros}, World Scientific, 2018
\bibitem{We2} L. WENG, Non-Abelian Zeta Function, Fokker-Planck Equation and Projectively Flat Connection, arXiv:1903.03604 
\bibitem{WZ} L. WENG, D.Zagier, Higher rank zeta functions for elliptic curves,
Proc. Natl. Acad. Sci. USA 117 (2020), no.9, 4546-4558
\bibitem{WZ2} L. WENG, D.Zagier, Higher rank zeta functions and $\SL_n$-zeta functions for curves,
Proc. Natl. Acad. Sci. USA 117 (2020), no.12, 6279-6281
\bibitem{Y} M.S. Ying, {\it Foundations of Quantum Programming}, Elsevier, 2016
\end{thebibliography}
\end{document}